\documentclass[reprint,prb,superscriptaddress,amsmath,amssymb,aps
]{revtex4-2}

\usepackage{bm}%
\usepackage[colorlinks=true,linkcolor=blue]{hyperref}%
\expandafter\ifx\csname package@font\endcsname\relax\else
 \expandafter\expandafter
 \expandafter\usepackage
 \expandafter\expandafter
 \expandafter{\csname package@font\endcsname}%
\fi
\usepackage{graphicx}
\usepackage{dcolumn}

\begin{document}

\title{Engineering Delocalization in Graphene Nanoribbons via Quasiperiodic Edges and Electronic Interactions}

\author{Diego B. Fonseca}
\affiliation{Departamento de F\'{\i}sica, Centro de Ci\^encias Exatas e da Natureza, Universidade Federal de Pernambuco, Recife - PE, 50670-901, Brazil}

\author{Anderson L. R. Barbosa}
\affiliation{Departamento de F\'{\i}sica, Universidade Federal Rural de Pernambuco, Recife - PE, 52171-900, Brazil}

\author{Luiz Felipe C. Pereira}
\affiliation{Departamento de F\'{\i}sica, Centro de Ci\^encias Exatas e da Natureza, Universidade Federal de Pernambuco, Recife - PE, 50670-901, Brazil}

\begin{abstract}
We investigate localization effects in zigzag graphene nanoribbons with quasiperiodic Fibonacci-type edge extensions, accounting for electron-electron interactions. 
We employ a tight-binding model that includes first- and third-nearest-neighbor hoppings, in which electronic interactions are treated within a self-consistent mean-field Hubbard approximation. 
Charge transport properties are calculated using the Landauer-B\"uttiker formalism. 
Our results reveal that the combination of quasiperiodic geometry and electronic interactions gives rise to nontrivial transport phenomena.
Specifically, the system exhibits three transport regimes: in the non-interacting case, we observe geometric localization. For weak interactions, the system shows a conductive regime with \textcolor{black}{transmission oscillations}, whose multiplicity increases with the Fibonacci generation order. 
In this regime, delocalization emerges from the interplay between geometry and interaction-induced correlations. 
Finally, for strong interactions, repulsion dominates, and the system returns to a localized state. 
Our results  demonstrate that quasiperiodic edge engineering, combined with electronic interaction control, offers a promising path to modulate transport in graphene nanoribbons.

\end{abstract}


\maketitle

\section{Introduction} 

The isolation of graphene initiated a sustained effort to explore two-dimensional materials, elevating them to a central role in condensed-matter physics and materials science due to their exceptional mechanical, optical, electronic, and magnetic characteristics [\onlinecite{ref1,RevModPhys.92.021003,PhysRevB.107.155432,Sierra_2021,Perkins_2024,D3NH00416C,D5NA00532A}]. Among these systems, graphene nanoribbons (GNRs) occupy a privileged position as quasi-one-dimensional platforms whose low-energy spectrum can be engineered through width, edge termination, and lateral functionalization.
Such control enables band-gap engineering, the stabilization of spin-polarized edge states, and the realization of topological electronic phases relevant for nanoelectronics and spintronics [\onlinecite{RevModPhys.92.021003,Perkins_2024,PhysRevMaterials.9.014203,PhysRevB.110.075421,PhysRevB.111.035411}].

The edge structure of GNRs -- whether armchair, zigzag, or more complex motifs -- critically governs their electronic and magnetic behavior. Advances in chemical synthesis now enable atomically precise GNRs with tailored edge geometries, overcoming limitations of top-down fabrication [\onlinecite{PhysRevMaterials.9.014203, xiang2025zigzag, houtsma2021atomically}]. Bottom-up approaches have yielded specialized nanostructures including edge-extended GNRs [\onlinecite{https://doi.org/10.1002/adma.202306311}] and porphyrin-functionalized zigzag ribbons [\onlinecite{xiang2025zigzag}], which introduce localized spin-polarized states, tunable band gaps, magnetic anisotropy, and topological phases. These developments expand GNRs' application potential in nanoelectronics and spintronics.

Electron-electron interactions further enrich GNR physics, particularly in confined geometries where on-site Coulomb repulsion can drive metal-insulator transitions, stabilize magnetic ordering, and reshape electronic spectra [\onlinecite{RevModPhys.81.109}]. 
Recent studies employing Hubbard mean-field models have explored how edge disorder influences transport in zigzag GNRs [\onlinecite{mukim2025edge}].
The strength of these interactions can be modulated through proximity screening effects, as demonstrated by controlled tuning of electron-electron scattering lengths in graphene via atomically thin gate dielectrics [\onlinecite{Kim2020control}].
While pristine graphene remains non-magnetic at half-filling due to the vanishing density of states at the Dirac point, edge-localized states in nanostructured ribbons exhibit strong correlation effects [\onlinecite{fujita1996peculiar, son2006half, doi:10.1021/acs.nanolett.9b04075}]. 

The interplay between engineered edges and interactions enables the emergence of correlation-driven metallic phases, spin-polarized currents, and nontrivial topological responses beyond non-interacting descriptions. \textcolor{black}{This interplay is particularly relevant in the context of Anderson localization, where disorder or geometric perturbations can exponentially suppress transmission with increasing system length, such that $T \propto e^{-L/\xi}$, where $\xi$ is the localization length [\onlinecite{PhysRevLett.42.673,PhysRev.109.1492}]. In GNRs, Anderson localization has been observed in samples with edge disorder [\onlinecite{PhysRevB.78.161407, PhysRevB.79.075407, 10.1063/1.4966173}], defect-induced [\onlinecite{doi:10.1021/nn401992q}], adatom doping [\onlinecite{PhysRevB.90.085425, PhysRevLett.113.246601}], and substrate-induced randomness [\onlinecite{PhysRevB.76.214204}]. Such interplay is further} exemplified by the non-monotonic effect of electron-electron interactions on Anderson localization, where a Hubbard interaction can suppress localization at weak coupling but reinforce it at strong coupling, as demonstrated for two-electron wavepackets in disordered chains [\onlinecite{DIAS201435}]. 
Similarly, spatial correlations in on-site energies or coupling disorder can suppress Anderson localization, leading to an extended state regime with significantly increased localization length, in contrast to the complete localization observed for uncorrelated disorder [\onlinecite{junior2023magnon, fonseca2024disordered,10.1063/5.0275322,BARBOSA2020114210}].

\textcolor{black}{Recent experimental advances have demonstrated remarkable control in the synthesis of GNRs with properties designed through edge engineering [\onlinecite{https://doi.org/10.1002/adma.202306311}]. In this work, edge-extended nanoribbons feature deliberate zigzag-edge modifications that alter local electronic states and magnetic behavior, enabling novel properties including localized spin-polarized edge states, topological characteristics, and tunable energy gaps.} Similarly, by incorporating transition-metal porphyrins into zigzag ribbons one can tune the chemical-electronic-magnetic interplay for high-performance quantum devices [\onlinecite{xiang2025zigzag}]. 
These approaches facilitate coherent spin manipulation while integrating quantum functionality with electronic properties for emerging technologies.

\begin{figure*}[htb]
    \centering
    \includegraphics[width=0.80\linewidth]{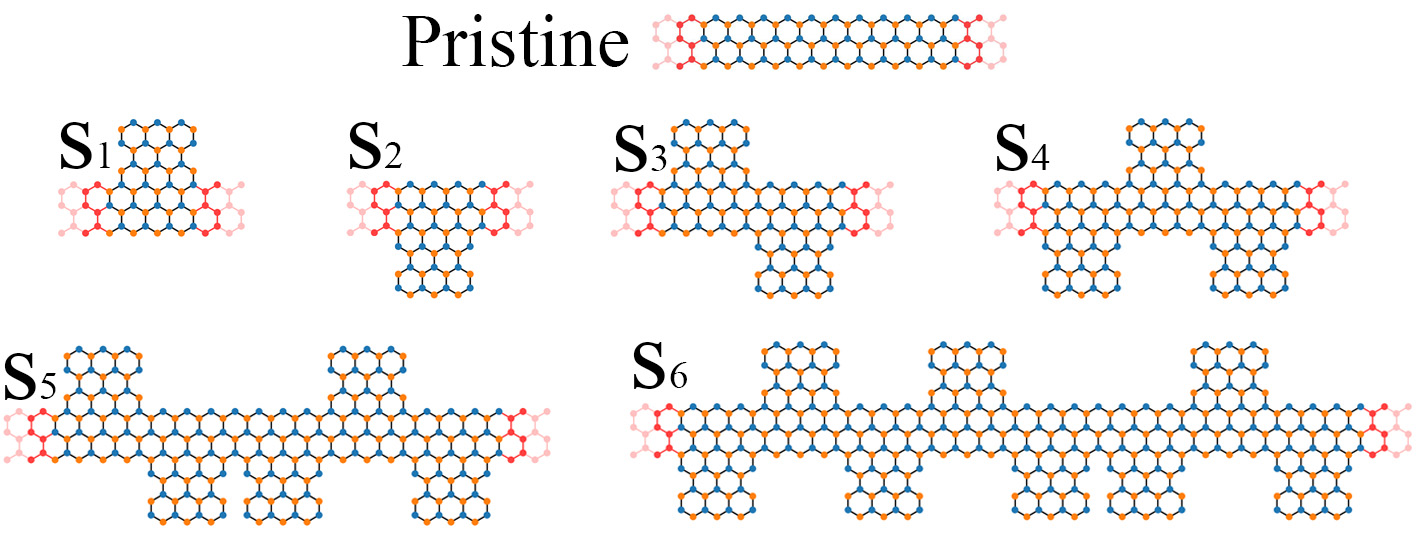}
    \caption{Illustration of a pristine zigzag nanoribbon, as well as the first six generations of Fibonacci edge-extended ribbons.}
    \label{fig:redes_fibo}
\end{figure*}

While most experimental research has focused on periodic GNR architectures with regular edge extensions or fused molecular units [\onlinecite{https://doi.org/10.1002/adma.202306311,xiang2025zigzag}], quasiperiodic patterns -- particularly Fibonacci sequences -- offer promising alternatives for transport engineering. Theoretical studies demonstrate that quasiperiodic modulations in hopping or edge structures can open conductance gaps, generate unconventional spectral features, and induce   \textcolor{black}{Anderson localization when a quasiperiodic scattering region is embedded between semi-infinite periodic leads [\onlinecite{da2022localization}]. This contrasts with the hallmark of  quasiperiodic order, which, as opposed to random disorder, is the emergence of critical states with multifractal characteristics, lying between the extremes of extended Bloch states and localized ones [\onlinecite{10.21468/SciPostPhys.6.4.050, PhysRevB.93.205153, PhysRevResearch.3.033257, PhysRevA.35.1467, PhysRevB.35.1020, PhysRevB.34.2041}].} These architectures offer finer control than random disorder, enabling richer physics than periodic systems, and creating versatile platforms to explore structural determinism, quantum interference, and electronic correlations.

\textcolor{black}{Motivated by the recent synthesis of zigzag GNRs with periodic edge modulations reported in Ref.  [\onlinecite{https://doi.org/10.1002/adma.202306311}],
in the present work, we explore the electronic transport properties of zigzag GNRs whose edges are modulated following the quasi-periodic Fibonacci sequence.
We explicitly consider electron–electron interactions via a self-consistent mean-field Hubbard approach.
Green’s function transport calculations are employed to study  electronic transport features in the zigzag GNRs such as conductance, local density of states, phase diagram and localization-delocalization-localization regimes. 
We demonstrate that quasiperiodic edge engineering leads to nontrivial conductance oscillations and delocalization regimes driven by the competition between geometry and electronic correlations.}

The paper is organized as follows. Section~\ref{sec:method} introduces the microscopic model and self-consistent computational procedure. In Section~\ref{sec:results}, we present transport results for different Fibonacci generations and stacked-block configurations, comparing non-interacting and interacting regimes. Section~\ref{sec:discussion} analyzes the underlying physical mechanisms, and Section~\ref{sec:conclusions} summarizes our findings and perspectives.

\section{Methodology}  \label{sec:method}

We investigate charge transport in Fibonacci edge-extended zigzag GNRs (FEE-ZGNRs) with electron-electron interaction. 
Each Fibonacci generation \(S_i\) is constructed by the recursion \(S_1=A\), \(S_2=B\) and \(S_i=S_{i-2}+S_{i-1}\) for \(i\ge3\), where the sum denotes the concatenation of the previous two generations, as illustrated in Figure \ref{fig:redes_fibo}.
The resulting \textcolor{black}{sequence} determines the positions of the edge extensions ($A$ or $B$) along the nanoribbon and is used to build stacked-block geometries explored further on.
\textcolor{black}{Note that the $S_3$ Fibonacci sequence is the zigzag GNR synthesized in Ref. [\onlinecite{https://doi.org/10.1002/adma.202306311}]. 
Therefore, the quasiperiodic GNRs  explored here are experimentally feasible and our predictions can be verified experimentally, at least in principle.}

We will now introduce the microscopic tight-binding model for graphene with electron-electron interactions and then we will describe the self-consistent method used. 

\begin{figure*}
  \centering
  \includegraphics[width=0.8\linewidth]{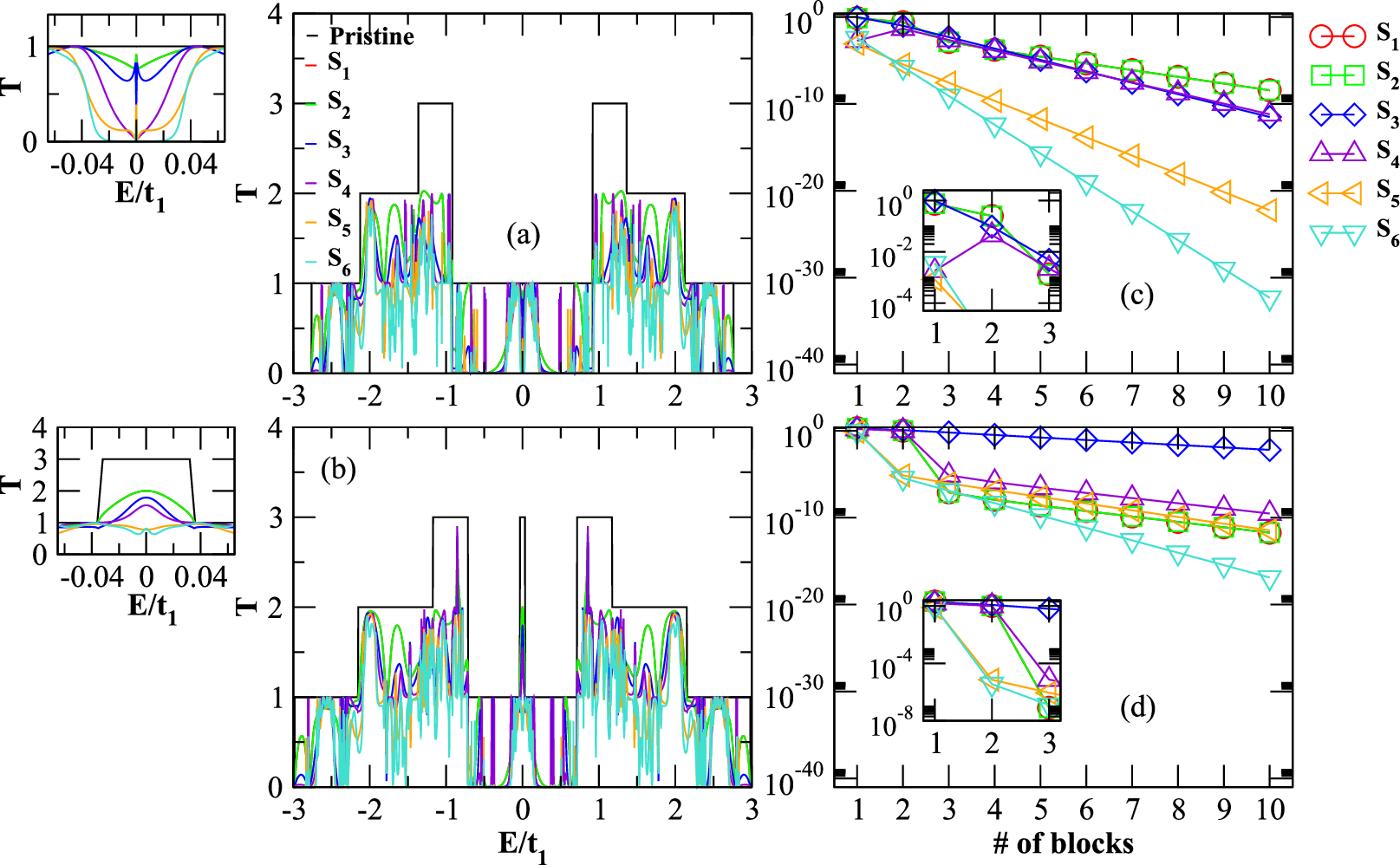}
  \caption{Main panels show charge transmission as a function of Fermi energy and number of blocks, for several generations of FEE-ZGNRs shown in Figure \ref{fig:redes_fibo}. 
  (a) Charge transmission considering only first-nearest-neighbors ($t_3 = 0$) and (b) with first- and third-nearest-neighbors ($t_3 = t_1/10$).
  Insets on the left-hand-side show the region around $E=0$.
  (c) Charge transmission at $E = 0$ considering only first-nearest-neighbors ($t_3 = 0$) and (d) with first- and third-nearest-neighbors ($t_3 = t_1/10$). Insets in (c) and (d) expand the region around 1,2  and 3 blocks.}
  \label{fig:T}
\end{figure*}

\subsection{Microscopic Model}
The charge transport through GNRs is characterized by the scattering matrix given by  [\onlinecite{datta}]
\begin{equation}
    S = 
    \begin{bmatrix}
       {r} &{t'}\\
        {t} &{r'}
    \end{bmatrix},
\end{equation}
where ${t}({t'})$ and ${r}({r'})$ are the transmission and reflection matrix blocks, respectively. 
The charge transmission coefficient can be calculated from the Landauer-B\"uttiker relation
\begin{equation}\label{eq:T}
    T(E)=\mathrm{Tr}\big[t(E)\,t^\dagger(E)\big]
\end{equation}
which is valid in the linear response regime and at low temperatures. Although originally derived for non-interacting systems, the Landauer-Büttiker formalism remains applicable to interacting electron systems: at zero temperature, energy conservation restricts transport to elastic single-electron processes, ensuring the validity of the formalism in the linear response regime [\onlinecite{meir1992landauer}]. This approach has been successfully implemented within first-principles schemes [\onlinecite{ferretti2005first, ferretti2005correlation}] and self-consistent mean-field frameworks [\onlinecite{rocha2006spin}]. While mean-field treatments provide approximate descriptions of electron-electron interactions, they enable the computational tractability of larger systems and have been recently applied to study GNRs with edge disorder and magnetism [\onlinecite{mukim2025edge}].

Numerical calculations of the transmission coefficient were performed with KWANT software [\onlinecite{kwant}], which is a Green’s function–based algorithm within the tight-binding approach.
\textcolor{black}{For our numerical experiments, we employed the tight-binding Hamiltonian for edge-extended zigzag graphene proposed in Ref.  [\onlinecite{https://doi.org/10.1002/adma.202306311}], which adheres to all relevant experimental conditions and includes mean-field electron-electron interactions. It is given by}
\begin{equation}\label{eq:H_mf}
    {H}_{MF} =
    - t_1 \sum_{\langle i,j \rangle} c_{i}^{\dagger} c_{j}
    - t_3 \sum_{\langle l,k \rangle} c_{l}^{\dagger} c_{k}
    + U \sum_{i} \langle n_{i} \rangle n_{i},
\end{equation}
where the indices $i$ and $j$ run over all lattice sites, and $\langle i,j \rangle$ denotes first nearest neighbors. The operators $c_{i}^\dagger$ and $c_{i}$ create and annihilate, respectively, an electron at site $i$. Similarly, the indices $l$ and $k$ run over all sites, but $\langle l,k \rangle$ refers to the third nearest neighbors.
The first term in $\hat{H}_{MF}$ describes the standard nearest-neighbor hopping between lattice sites, with hopping amplitude $t_1$, typically equal to 2.7 eV [\onlinecite{RevModPhys.81.109}]. The second term accounts for third nearest-neighbor hopping processes, where we consider $t_3 = t_1/10$ in this work, consistent with first-principles parametrizations for systems with similar structure [\onlinecite{https://doi.org/10.1002/adma.202306311}]. \textcolor{black}{ Note that the second-neighbor hopping $t_2$ was set to zero to maintain consistency with the experiment and the tight-binding model reported in Ref. [\onlinecite{https://doi.org/10.1002/adma.202306311}]. Including a non-zero $t_2$ would primarily break electron-hole symmetry at $E = 0$, which was not experimentally observed, and shift the band structure rigidly in energy bands [\onlinecite{RevModPhys.81.109}], resulting in a rigid shift of the transmission spectrum without altering the relative trends or qualitative conclusions reported here.} Finally, the third term represents the on-site Hubbard interaction in the mean-field approximation. Here, $U$ corresponds to the strength of the repulsive e-e interaction, $n_{i} = c_{i}^{\dagger} c_{i}$ is the number operator, and $\langle n_{i} \rangle$ is the mean occupation at site $i$.

It is important to stress that Eq.~\eqref{eq:H_mf} originates from the standard Hubbard Hamiltonian with first- and third-nearest-neighbor hoppings,
\begin{equation}
\begin{split}
H ={}& -t_1\sum_{\langle i,j\rangle,\sigma} c_{i\sigma}^\dagger c_{j\sigma}
- t_3\sum_{\langle l,k\rangle,\sigma} c_{l\sigma}^\dagger c_{k\sigma} \\
&+ U\sum_{i,\sigma} \langle n_{i\sigma} \rangle n_{i-\sigma},
\end{split}
\end{equation}
but is here simplified by imposing a spin-symmetric reduction. Specifically, we assume
\begin{equation}
\langle n_{i\uparrow}\rangle = \langle n_{i\downarrow}\rangle = \langle n_{i}\rangle \quad \forall i,
\end{equation}
so that the per-spin occupation is identical at every site. Under mean-field decoupling, this constraint transforms the interaction term into an effective on-site potential proportional to the local charge density. Such a procedure reduces the number of degrees of freedom while still retaining charge-redistribution effects, thereby allowing us to focus on the interplay between quasiperiodic geometry and electron–electron interaction. We deliberately restrict our self-consistent calculations to this spin-symmetric solution, since our goal is to investigate charge transport rather than spin polarization. This enables simulations of considerably larger systems, which is an essential requirement for resolving quasiperiodic transport trends.

\subsection{Self-Consistency} \label{subsec:SC}

We employ a self-consistent mean-field treatment of the Hubbard term. Despite its limitations regarding dynamical correlations, this approach has been shown to reproduce DFT trends for GNRs [\onlinecite{mukim2025edge, Yazyev_2010}] while enabling the simulation of much larger quasiperiodic structures, which provides a crucial advantage for our study.
The mean-field Hamiltonian is not known explicitly; instead, it depends functionally on the charge density $\langle n_i \rangle$. This is the standard situation in mean-field electronic structure theory calculations [\onlinecite{rocha2006spin}]. In equilibrium the lesser Green's function relates to the retarded component via
\begin{equation}
G^<(E) = -2i\,\mathrm{Im}\,G^R(E)\,f(E-\mu),
\end{equation}
and the per-site occupation is obtained by integrating the LDOS up to the Fermi energy:
\begin{equation}
\begin{split}
\langle n_i \rangle {}& \;=\; \frac{1}{2\pi i}\int_{-\infty}^{\infty}\big[G^<(E)\big]_{ii}\,dE \\
{}& \;=\; \frac{1}{\pi}\int_{-\infty}^{E_F}\mathrm{Im}\big[G^R(E)\big]_{ii}\,dE .
\label{eq:rho_from_ldos}
\end{split}
\end{equation}

In our implementation, we compute $\langle n_i \rangle$ by numerical integration of the LDOS. Specifically, the LDOS is evaluated on a uniform energy grid and the integral in Eq. ~\eqref{eq:rho_from_ldos} is approximated with the trapezoidal rule. \textcolor{black}{To validate this procedure, we compared two independent implementations available in KWANT: a Green's function approach, in which a small imaginary regularization $E \to E + i\eta$ (with $\eta$ ranging from $10^{-6}t_1$ to $10^{-8}t_1$) is added to the energy, and a direct evaluation of the LDOS from the scattering wave functions, which requires no broadening parameter. Both methods yield qualitatively equivalent results, confirming robustness regardless of the numerical implementation.} The typical energy window and discretization used in the reported runs are
$$
E\in[-3t_1,\,E_F],\qquad \Delta E = 0.01\,t_1.
$$

The self-consistent loop implemented in the code proceeds as follows:
\begin{enumerate}
  \item Initialize the per-site densities with a uniform half-filling seed: $\langle n_i \rangle^{(0)}=0.5$.
  \item Build the device Hamiltonian \(H_{MF}[\{\langle n_i \rangle^{(k)}\}]\) by assigning on-site potentials \(U\,\langle n_i \rangle^{(k)}\).
  
  \item Compute the LDOS on the energy grid; integrate the LDOS to obtain \(\langle n_i \rangle^{\text{new}}\).
  
  \item Form the residual $r^{(k)}=\langle n_i \rangle^{\text{new}}-\langle n_i \rangle^{(k)}$ and update the density using an Anderson mixing routine [\onlinecite{NOVAK2023108865, walker2011anderson, zhang2020globally}] with history length 5 and damping parameter $\alpha=0.1$. 
  
  \item Repeat steps 2–4 until the relative convergence criterion
      \begin{equation}
           \frac{\|\langle n_i \rangle^{(k+1)}-\langle n_i \rangle^{(k)}\|}{\|\langle n_i \rangle^{(k)}\|} < 10^{-6}
      \end{equation}
  is met or until a maximum of \(25{,}000\) iterations is reached.
\end{enumerate}

\textcolor{black}{Note that this criterion monitors the relative change in the on-site electron density between successive iterations, and is independent of any energy regularization parameter; the chosen tolerance of $10^{-6}$ was sufficient to achieve self-consistency in the present calculations, concordant with standard practice in self-consistent transport methodologies [\onlinecite{NOVAK2023108865}].}
Once convergence is reached for a given configuration (Fibonacci generation, number of stacked blocks \(N\), Hubbard \(U\), and third-neighbor toggling), the observables of interest are extracted as scattering matrix, transmission, and density.

\section{Results}\label{sec:results}

We will split this section into two parts. The first part presents the results for non-interacting FEE-ZGNRs, such that $U = 0$ in Eq.~\eqref{eq:H_mf}. Meanwhile, the second part addresses interacting FEE-ZGNRs, in which $U > 0$.

\begin{figure*}
\centering
\includegraphics[width=0.9\linewidth]{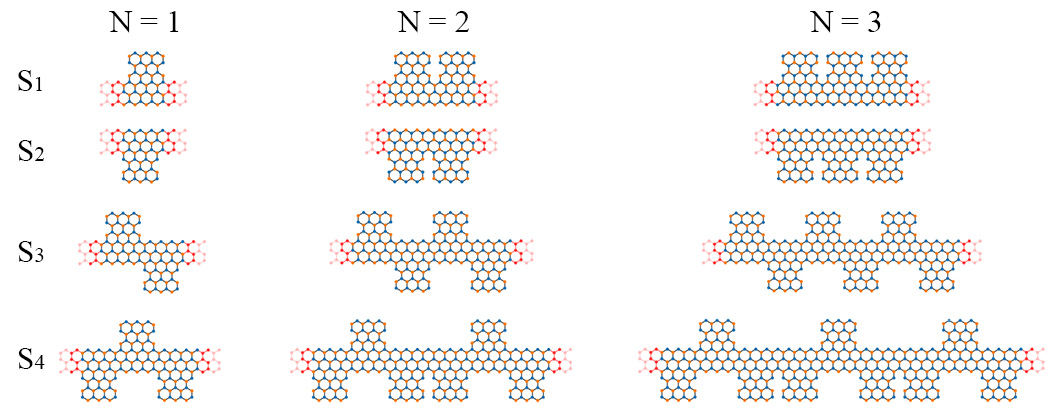}
\caption{Illustration of the first four generations of Fibonacci edge-extended zigzag graphene nanoribbons with multiple block repetitions: $N=1,2$ and $3$.}
\label{fig:redes_blocos}
\end{figure*}

\subsection{Non-interacting FEE-ZGNRs}

Figures \ref{fig:T}(a) and \ref{fig:T}(b) show charge transmission, Eq. ~\eqref{eq:T}, as a function of Fermi energy for the first six generations of FEE-ZGNRs shown in Figure \ref{fig:redes_fibo} for $U=0$, alongside a pristine ribbon for reference. 
In Figure \ref{fig:T}(a), where only nearest-neighbor hoppings are considered, i.e. \(t_3 =0\), the Fibonacci generations exhibit a clear reduction in transmission relative to the pristine case, with higher-order generations displaying increasingly pronounced fluctuations. Meanwhile, in Figure \ref{fig:T}(b), where third-nearest-neighbor interactions are included, a crucial change emerges:  additional propagating channels open at $E=0$, enabling the Fibonacci generations to transmit more effectively. Although this enhancement gradually diminishes for higher generations, the overall behavior away from $E=0$ remains qualitatively similar to the nearest-neighbor case. \textcolor{black}{Notably, the transmission fluctuations observed across all generations, in both hopping models, are multifractal as confirmed by a Multifractal Detrended Fluctuation Analysis (MF-DFA), which we present in the Appendix, reflecting the quasiperiodic criticality inherent to Fibonacci chains [\onlinecite{10.21468/SciPostPhys.6.4.050, PhysRevB.93.205153, PhysRevResearch.3.033257, PhysRevA.35.1467, PhysRevB.35.1020, PhysRevB.34.2041}].}

For further analysis, we constructed nanoribbons by stacking blocks of a given Fibonacci generation, as illustrated in Figure \ref{fig:redes_blocos}. Each generation $S_i$ is regarded as a fundamental block (or unit), such that $N=1$ corresponds to the smallest unit itself, while $N=2$ and $N=3$ correspond to two and three identical blocks joined sequentially, respectively. We stack blocks up to $N = 10$, and investigate localization effects induced by the Fibonacci edge extensions by analyzing the electronic transmission at $E=0$.
Figures \ref{fig:T}(c) and \ref{fig:T}(d) present the charge transmission as a function of the number of stacked blocks for different generations of FEE-ZGNRs. In Figure \ref{fig:T}(c), where only nearest-neighbor hopping is considered, we observe a pronounced exponential \textcolor{black}{suppression of transmission, i.e. Anderson localization, for all generations as the number of blocks increases, which becomes stronger with higher-order generations}. As shown in the inset for generations of order 1 to 4, the decay rate (slope on a log scale) changes between the first two segments. 
Beyond $N=3$, the slope stabilizes to a constant value. In contrast, the 5th- and 6th-order generations maintain the same slope throughout. 

With the inclusion of third-nearest-neighbor interactions, Figure \ref{fig:T}(d) shows that\textcolor{black}{, despite the additional hopping, increasing Fibonacci generation enhances confinement and suppresses the transmission channels.} However, the decay is significantly slower compared to the first-nearest-neighbor model, Figure \ref{fig:T}(c). This overall slower \textcolor{black}{electronic} localization for all generations is consistent with Figure \ref{fig:T}(b), where the additional third-nearest-neighbor interactions open more propagating channels at $E=0$. Among the different generations, the third generation localizes much slower than the others. This enhanced robustness against localization can be attributed to its particular structure, since it is the only generation with an equal number of alternating A- and B-type edge extensions, as illustrated by Figure \ref{fig:redes_blocos}. Unlike the other generations, it maintains a constant decay rate from the beginning, as shown in the inset of Fig. \ref{fig:T}(d).

\begin{figure}
    \centering
    \includegraphics[width=0.9\linewidth]{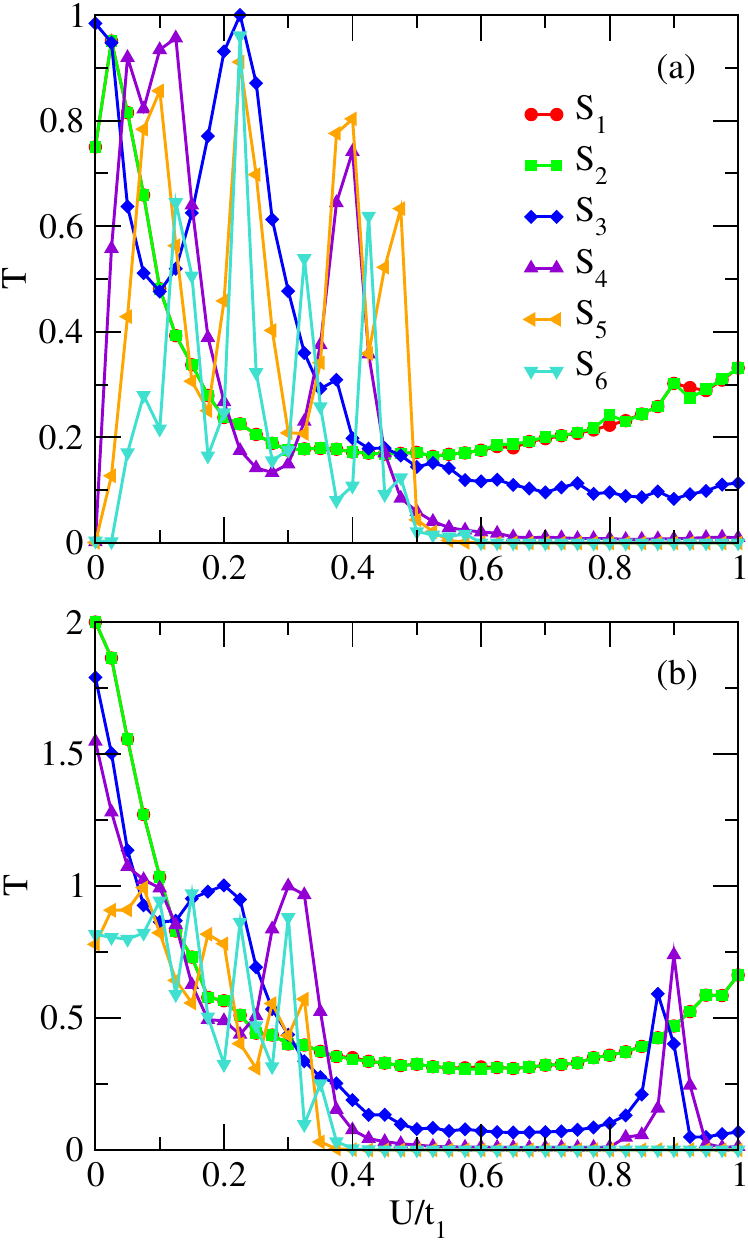}
    \caption{Charge transmission  through the different generations of FEE-ZGNRs depicted in Figure \ref{fig:redes_fibo}, as a function of Hubbard interaction U at energy $E = 0$. (a) Only first-nearest-neighbors ($t_3 = 0$) and (b) first- and third-nearest-neighbors ($t_3 = t_1/10$).}
    \label{fig:T_U}
\end{figure}

The inclusion of electron-electron interaction, modeled here within the Hubbard approach, strongly affects charge transport in FEE-ZGNRs. For a direct comparison with the non-interacting case, we focus again on the transmission at $E=0$. 
Figures \ref{fig:T_U}(a) and \ref{fig:T_U}(b) show the transmission through different generations of FEE-ZGNRs, depicted in Figure \ref{fig:redes_fibo}, as a function of the interaction strength $U$ for the cases with only nearest-neighbor hopping and with both nearest- and third-nearest-neighbor hopping, respectively. In both cases, \textcolor{black}{the number of oscillations in the transmission increases systematically with the Fibonacci generation: $S_1$ and $S_2$ show a single oscillation, $S_3$ shows two, $S_4$ three, $S_5$ four, and $S_6$ five. Lower-order generations thus present weaker oscillatory behavior, while higher-order generations exhibit increasingly pronounced oscillations.}

In addition, there is a value of the electronic interaction beyond which transmission is strongly suppressed. For nearest-neighbor hopping only, this suppression occurs at $U \approx 0.5t_1$, while with third-nearest-neighbor hopping it shifts to $U \approx 0.4t_1$. In the third-nearest-neighbor case, $S_3$ and $S_4$ also exhibit \textcolor{black}{an oscillation} even after the initial transmission reduction.

\begin{figure*}
    \centering
    \includegraphics[width=0.95\linewidth]{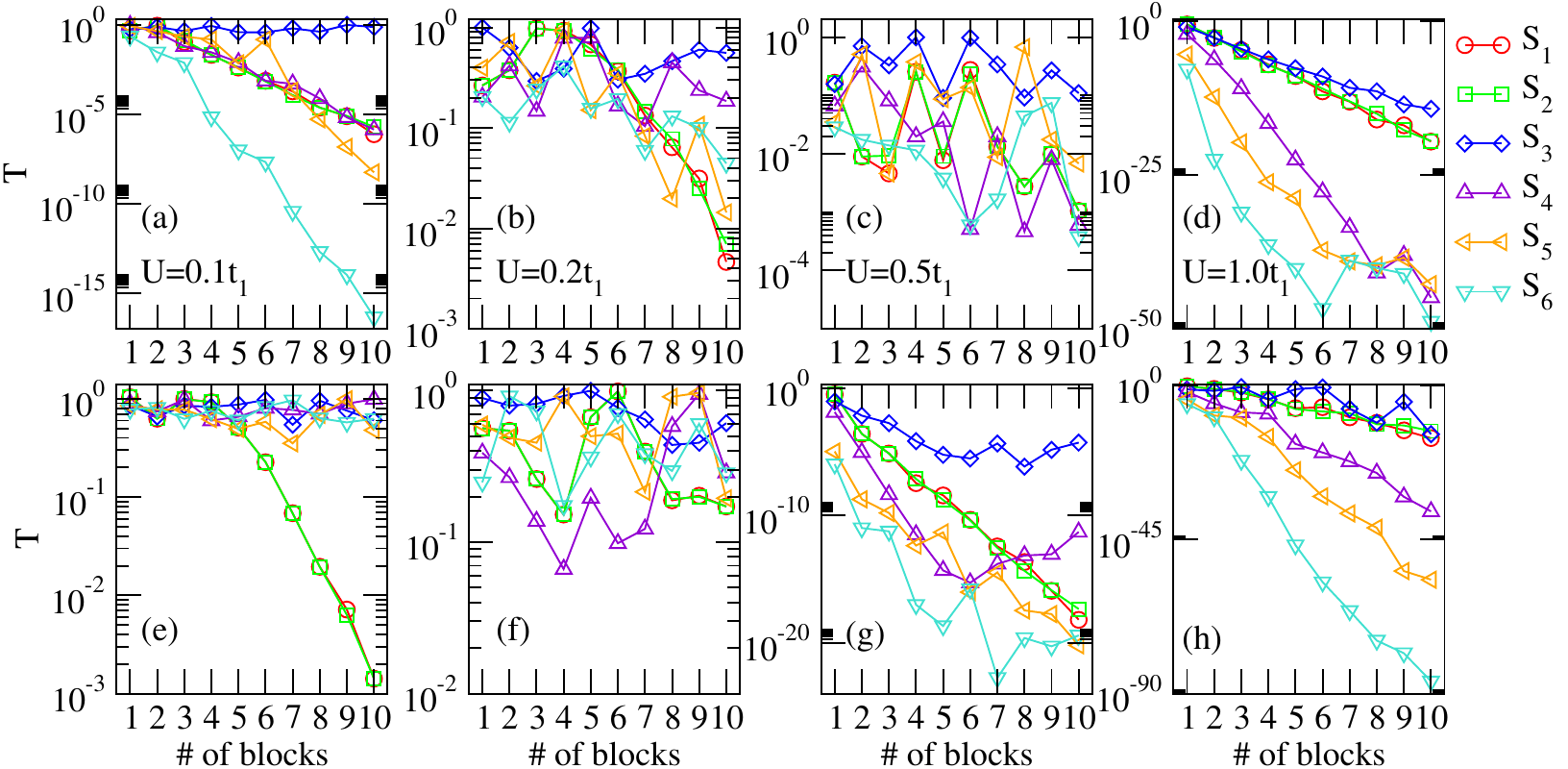}
    \caption{Charge transmission (symbols) through the different generations of FEE-ZGNRs depicted in Figure \ref{fig:redes_blocos}, as a function of the number of blocks at energy $E = 0$ and Hubbard interaction $U = 0.1t_1$, $0.2t_1$, $0.5t_1$, and $1.0t_1$ in each column from left to right. For (a-d) only first-nearest-neighbors are considered ($t_3 = 0$), while in (e-h) first- and third-nearest-neighbors ($t_3 = t_1/10$) are included. Solid lines are guides to the eyes.}
    \label{fig:T_N_U}
\end{figure*}

To further probe localization in the interacting regime, we analyzed the transmission as a function of the number of stacked blocks for different generations of FEE-ZGNRs, represented in Figure \ref{fig:redes_blocos}. We consider fixed finite values of $U$ as shown in Figure \ref{fig:T_N_U}, which is then equivalent to Figure \ref{fig:T} for the non-interacting case. 

\begin{figure}
    \centering
    \includegraphics[width=0.95\linewidth]{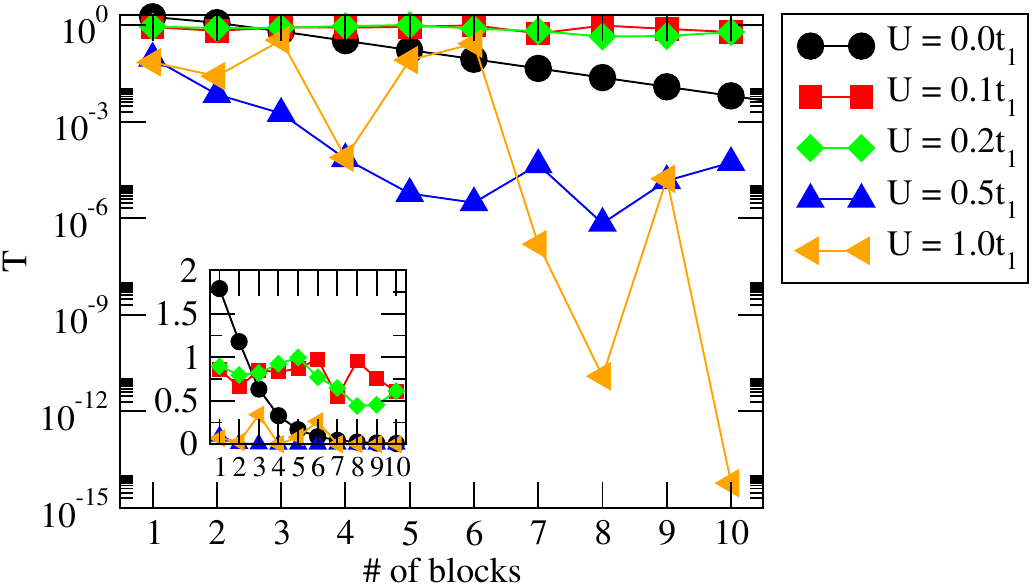}
    \caption{Charge transmission (symbols) through the periodic generations $S_3$ of FEE-ZGNRs depicted in Figure \ref{fig:redes_blocos}, as a function of the number of blocks at energy $E = 0$, \(t_3=t_1/10\) and Hubbard interaction $U = 0.0$, $0.1t_1$, $0.2t_1$, $0.5t_1$, and $1.0t_1$. Solid lines are guides to the eyes.}
    \label{fig:S3_T_N}
\end{figure}

Let us begin with a weak interaction $U = 0.1t_1$, shown in the first column of Figure \ref{fig:T_N_U}. For the nearest-neighbor case, Figure \ref{fig:T_N_U}(a), most generations display clear localization with increasing system size. The only exception is $S_3$, which maintains a finite transmission even at larger lengths, indicating enhanced robustness against localization, i.e., delocalization. Compared to the non-interacting case, Figure \ref{fig:T}(c), the overall decay is slower, reflecting the influence of electron-electron interaction-induced correlations. In the third-nearest-neighbor case, Figure \ref{fig:T_N_U}(e), we see that $S_1$ and $S_2$ still show a tendency toward localization with increasing block size, but the following  generations, $S_3$, $S_4$, $S_5$, and $S_6$, delocalize.

For $U = 0.2t_1$, the second column of Figure \ref{fig:T_N_U}, the nearest-neighbor case  Figure \ref{fig:T_N_U}(b) exhibits a marked change: while $S_1$ and $S_2$ still tend toward localization beyond about eight blocks, higher generations fluctuate but remain delocalized. With the inclusion of third-nearest-neighbor hopping, Figure \ref{fig:T_N_U}(f), all generations remain delocalized, although their transmission also shows oscillations.

At intermediate interaction $U = 0.5t_1$, the third column of Figure \ref{fig:T_N_U}, localization reemerges. In the nearest-neighbor case, the transmission alternates between localized and extended regimes as the number of blocks increases, while $S_3$ remains resistant to localization. With third-nearest-neighbor hopping, however, all generations return to a localized regime, although $S_3$ once again localizes much more slowly than the others. These two behaviors are consistent with Figure \ref{fig:T_U}: in the nearest-neighbor case, the observed intermittency arises because $U = 0.5t_1$ lies near the transition between conducting and insulating regimes; in the third-nearest-neighbor case, $U = 0.5t_1$ falls within the localized regime where transmission is already suppressed.

\textcolor{black}{For the strong interaction case $U = 1.0t_1$, the last column of Figure \ref{fig:T_N_U} shows that both  nearest- and third-nearest-neighbor cases exhibit robust localization across all generations. Yet, in both situations, $S_3$ consistently localizes more slowly, confirming its anomalous behavior across the entire range of interaction strengths. Finally, Figure \ref{fig:S3_T_N} shows \(T(N)\) through the periodic generation $S_3$ as a function of the number of blocks keeping fixed  $E = 0$, and \(t_3=t_1/10\) for different values of \(U\). For \(U=0.0\), we observe a localization regime, followed by delocalization for the weak interactions \(U=0.1t_1\) and \(U=0.2t_1\), reflecting the interaction-induced correlation. For strong interactions \(U=0.5t_1\) and \(U=1.0t_1\), the localization regime is recovered because electronic repulsion  dominates the transport.}

\section{Discussion}\label{sec:discussion}

Our results demonstrate that the combination of Fibonacci edge extension geometry and electron-electron interaction leads to non-trivial transport phenomena. Specifically, we observe three main behavior regimes: (i) in the non-interacting limit ($U=0$), FEE-ZGNRs exhibit a strong localization tendency when stacked into blocks as shown in Figure \ref{fig:T}(c-d). \textcolor{black}{This behavior is consistent with Anderson localization driven by destructive interference induced by quasiperiodic edge geometry in contact with semi-infinite leads, favoring exponential decay rather than critical behavior, in agreement with [\onlinecite{da2022localization}]}; (ii) for small but finite interactions, a regime of enhanced transmission emerges, termed here the \emph{delocalized regime}, accompanied by transmission \textcolor{black}{oscillations} modulated by $U$\textcolor{black}{. A pronounced transmission suppression occurs around $U \approx 0.5t_1$ for nearest-neighbor hopping only, shifting to $U \approx 0.4t_1$ when $t_3$ is included, as shown in Figure \ref{fig:T_U}}; and (iii) for strong interactions, the system returns to a localized regime, as expected from intensified electronic repulsion. This sequence of localized \(\to\) delocalized \(\to\) localized with increasing $U$ constitutes the core finding of this work: geometry and electronic correlations compete in a tunable manner, enabling a mechanism to switch between distinct transport regimes.

\begin{figure*}
    \centering
    \includegraphics[width=\linewidth]{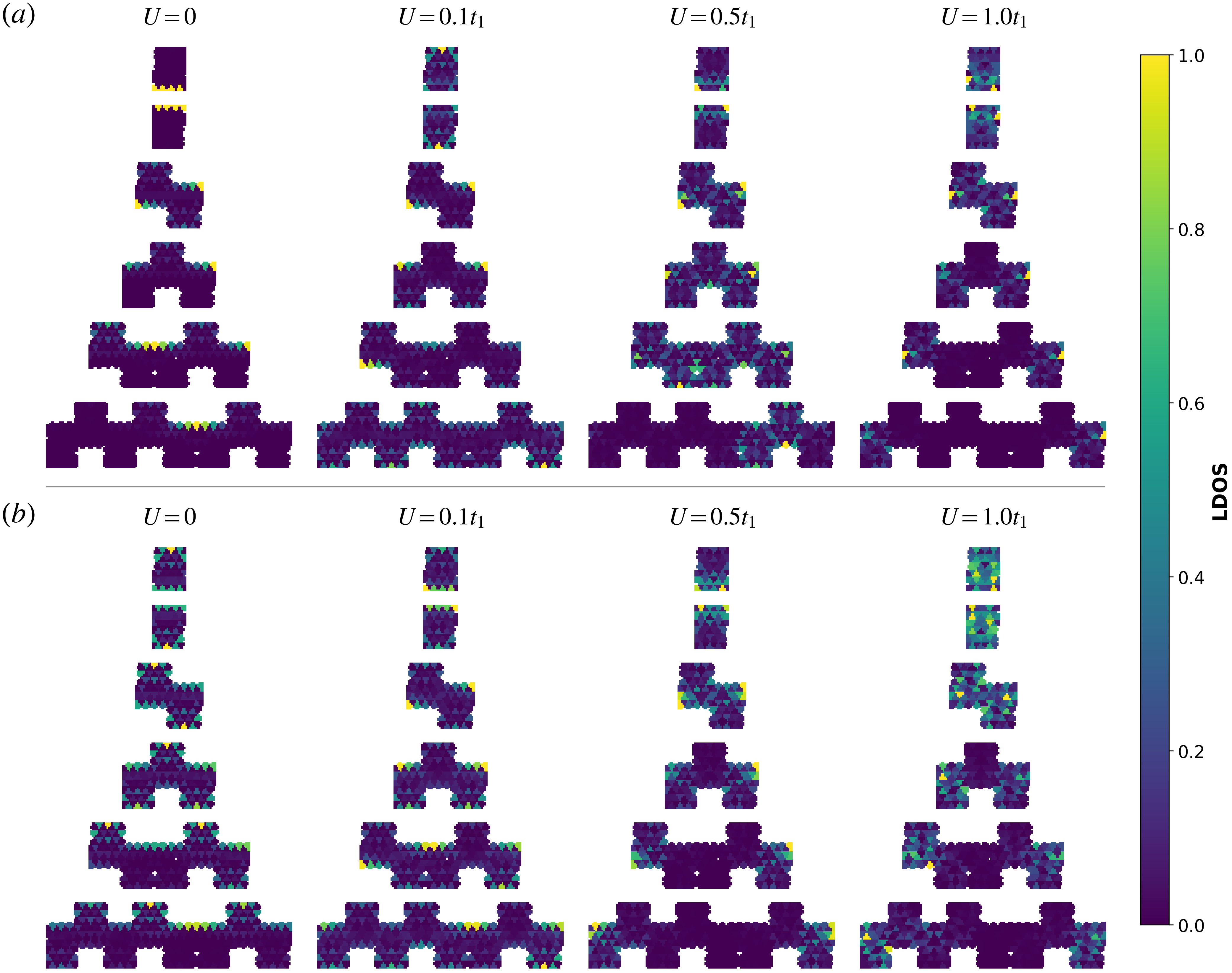}
    \caption{Normalized LDOS of FEE-ZGNRs for Fibonacci generations $S_1$ to $S_6$ (top to bottom) for each interaction strength $U = 0, 0.1t_1, 0.5t_1,\text{ and }1.0t_1$. (a) First-nearest-neighbor hopping only ($t_3 = 0$) and (b) first- and third-nearest-neighbor hopping ($t_3 = t_1/10$).}
    \label{fig:ldos}
\end{figure*}

\begin{figure}
    \centering
    \includegraphics[width=\linewidth]{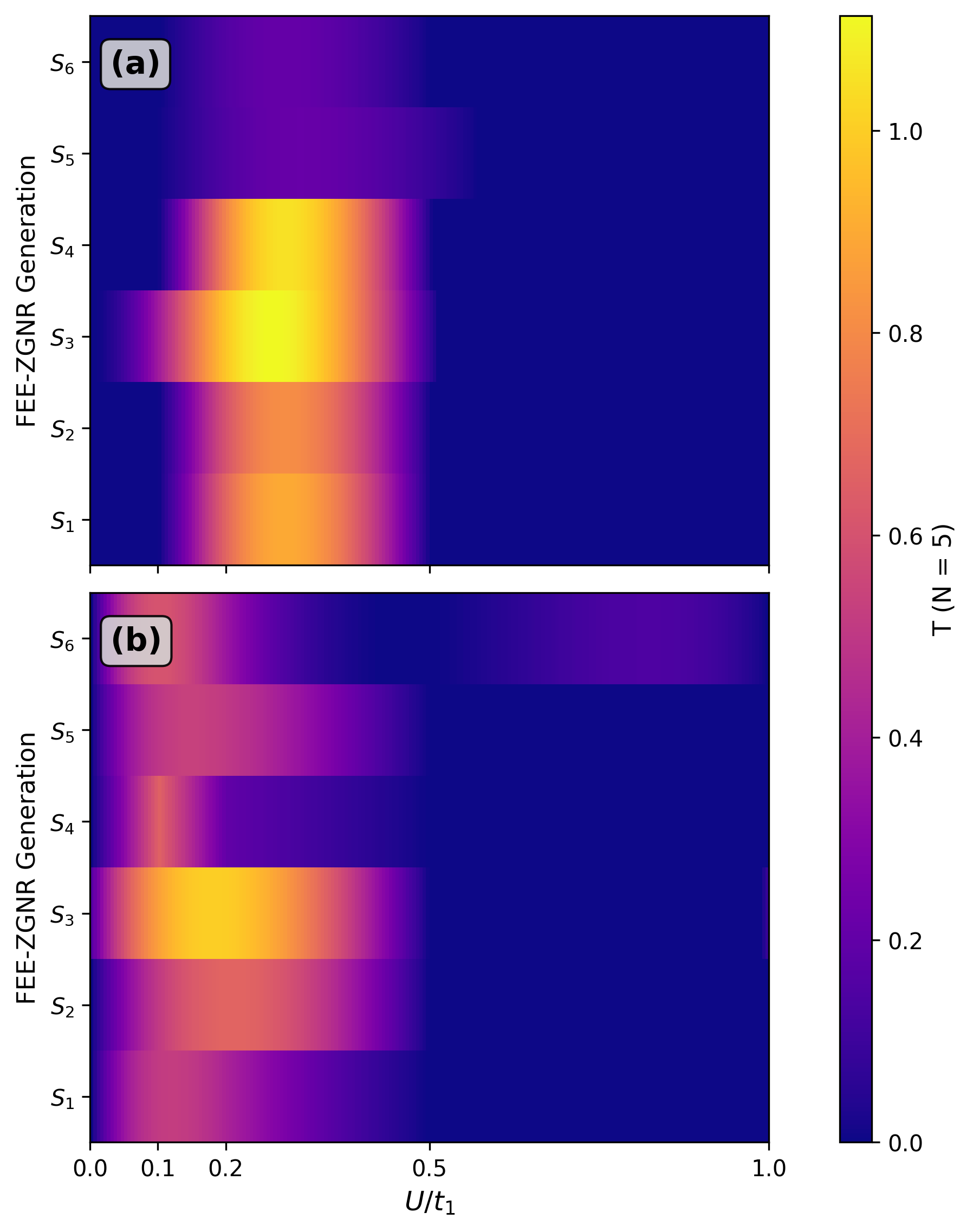}
    \caption{Heat map of charge transmission at $E = 0$ for $N = 5$ stacked blocks as a function of Fibonacci generation $S_i$ and Hubbard interaction strength $U$. (a) First-nearest-neighbor hopping only ($t_3 = 0$) and (b) first- and third-nearest-neighbor hopping ($t_3 = t_1/10$). Lighter (darker) regions correspond to finite (suppressed) transmission, identifying the delocalized (localized) regime.}
    \label{fig:transport_diagram}
\end{figure}

We interpret the transmission suppression observed in Figure \ref{fig:T_U} as the outcome of a competition between two effects induced by the electron-electron interaction: moderate interaction enhances correlations, giving rise to \textcolor{black}{oscillatory} transport, while strong interaction reinforces electron repulsion, which hinders conduction and ultimately drives the system toward localization.

The effect of including third-nearest-neighbor hopping is evident in our calculations: $t_3>0$ opens new transmission channels, as shown in Figure \ref{fig:T}(b), precisely at  $E=0$ where most analyses were performed. Consequently, the $T(U)$ curves and the $T$ dependence on the number of blocks exhibit a slower decay in the delocalized regime when $t_3$ is present, as observed in Figure \ref{fig:T_U}.

In fact, the introduction of electronic interactions via the mean-field Hubbard approximation reveals an even more complex behavior. We observe that moderate values of $U$ can effectively suppress the localization induced by quasiperiodicity, creating extended transport regimes observed in Figures \ref{fig:T_U} and \ref{fig:T_N_U}. This observation, that moderate interactions suppress geometric localization and create a conductive regime, is consistent with theoretical demonstrations in two-body models in periodic potentials showing that repulsion can induce delocalization [\onlinecite{vidal2000interaction}]. 
In that mechanism, the Hubbard term $U$ acts upon degenerate or near-degenerate states, redistributing the ground-state wave function and consequently increasing the single-particle localization length, which thus favors the appearance of extended regimes [\onlinecite{gulacsi2008delocalization}]. In our mean-field results, this corresponds to electronic correlation–induced delocalization, in which electronic repulsion, rather than simply hindering transport, reorganizes the electron density to \textcolor{black}{facilitate} transmission \textcolor{black}{with characteristic oscillations at specific $U$ values. Crucially, within these oscillatory \(U\) intervals, the system resists exponential suppression of transmission with increasing size, signaling the onset of delocalization. Notably,} the number of \textcolor{black}{these oscillations} increases systematically with the Fibonacci generation—i.e., higher-order generations exhibit richer \textcolor{black}{oscillatory} structures. This reinforces the idea that geometric complexity (the number and arrangement of A and B edge extensions) increases the spectrum of available electronic states that can be selectively accessed through correlation-driven charge redistribution.

\textcolor{black}{To further analyze our findings, we examine the local density of states (LDOS) for FEE-ZGNRs, as shown in Fig.~\ref{fig:ldos}. 
In order to examine all situations on an equal footing, we normalize the LDOS in each structure.  
We then consider two scenarios: one with only first-nearest-neighbors (panel a) and another with both first- and third-nearest-neighbors (panel b). 
In the non-interacting case (where \(U = 0\) and \(t_3 = 0\)), as illustrated in Fig.~\ref{fig:ldos}(a), the normalized LDOS indicates that for most generations exhibiting an imbalance between A- and B-type edge extensions: specifically \(S_1, S_2, S_4, S_5,\text{ and } S_6\),  electronic states tend to accumulate on the side of the ribbon with fewer edge defects. 
This finding suggests that conduction predominantly occurs along that side. }

\textcolor{black}{ When we introduce third-nearest-neighbor hopping, as shown in Fig.~\ref{fig:ldos}(b), the normalized LDOS becomes more evenly distributed across the sample. 
This observation aligns with the slower localization noted in Fig.~\ref{fig:T}(d). 
However, it can be noted that it is still insufficient to produce extended states. 
On the other hand, for the periodic generation \(S_3\), which features a balanced alternation of A-type and B-type extensions, we observe that  this combination promotes a redistribution of states throughout the entire nanoribbon, allowing for conduction along both sides and reducing the rate of localization. 
In contrast, \(S_4\) under the same conditions retains states localized on the side with fewer defects, even though the presence of third-nearest-neighbor hopping promotes a more homogeneous LDOS. 
This spatial difference in the LDOS is sufficient to explains why \(S_3\) and \(S_4\) yield nearly identical \(T(N)\) curves in the first-nearest-neighbol model: in both cases, the states are confined to the edges of the ribbon. 
However, when third-nearest neighbors are included they diverge significantly, with \(S_3\) demonstrating a greater resistance to localization.}

\textcolor{black}{For the interacting case \(U\neq0\), the LDOS in Fig.~\ref{fig:ldos}(a) reveals two qualitatively distinct localization mechanisms. 
At weak to intermediate interactions ($U = 0.1t_1$ and $U = 0.5t_1$), electron-electron correlations distribute the states more broadly across the nanoribbon, leading to extended states and facilitating transmission, which reflects the correlation-induced delocalization discussed above. 
At strong interaction ($U = 1.0t_1$), repulsion dominates and the states become confined near the leads, in contrast to the geometric localization at $U = 0$ where states accumulate at the ribbon edges. 
This spatial distinction in the LDOS provides a clear microscopic signature of two different localization mechanisms: one driven by quasiperiodic geometry, the other by strong electronic repulsion.
With third-nearest-neighbor hopping, Fig.~\ref{fig:ldos}(b), extended states persist for $U = 0.1t_1$, while for $U = 0.5t_1$ and $U = 1.0t_1$ the states are again localized near the leads, consistent with the earlier onset of the localized regime at $U \approx 0.4t_1$ observed in Fig.~\ref{fig:T_U}. 
This state redistribution thus provides a direct microscopic link to the transport behavior: the $U$ values at which extended states appear in the LDOS correspond precisely to those where $T(U)$ exhibits finite oscillatory transmission in Fig.~\ref{fig:T_U}, and where $T(N)$ resists exponential suppression in Fig.~\ref{fig:T_N_U}.}

\textcolor{black}{Figure~\ref{fig:transport_diagram} presents an additional perspective on the three transport regimes, displaying a heat map of transmission at \(E=0\) for \(N=5\) stacked blocks. 
This map illustrates how transmission varies as a function of Fibonacci generation \(S_i\) and interaction strength \(U\). 
In this context, lighter regions indicate finite, non-decaying transmission, corresponding to the delocalized regime, while darker areas indicate exponentially suppressed transmission.
In the first-nearest-neighbor case, shown in Fig.~\ref{fig:transport_diagram}(a), the delocalized regime is most prominent between \(U \approx 0.2t_1\) and \(0.5t_1\). 
Here, the higher generations \(S_5\) and \(S_6\) display somewhat weaker intensity. 
In contrast, when considering third-nearest-neighbor hopping, as depicted in Fig.~\ref{fig:transport_diagram}(b), the delocalized window shifts slightly to lower values, approximately \(U \approx 0.1t_1\) to \(0.4t_1\). 
This shift reflects the reduced critical interaction strength introduced by the additional hopping channels.
The periodic generation \(S_3\) shows the greatest resilience to this shift, consistent with its geometrically balanced edge structure, which provides enhanced stability across different interaction regimes. 
Overall, this diagram serves as a compact, unified visualization of the transition from localized $\to$ delocalized $\to$ localized transmission as a function of \(U\) across all the studied generations.}

Finally, we emphasize that identifying the $U$ intervals where conduction is enhanced, and where \textcolor{black}{oscillatory features of increasing complexity arise with the Fibonacci generation, indicates a practical path for transport engineering.
In principle, geometrical tuning of the edge extension sequence (and potentially control of $U$ via substrate or doping) allows modulation between localized and extended regimes, which translates into a tuning of the electronic transmission.}

\section{Conclusions} \label{sec:conclusions}

In summary, we have demonstrated that combining Fibonacci-sequence edge extensions with electron-electron interactions, treated within a mean-field framework, yields rich and controllable transport behavior. 
Starting from a localized regime at $U=0$, the inclusion of small—but finite—interactions can promote a delocalization window accompanied by \textcolor{black}{oscillatory transmission features whose complexity} increases with the Fibonacci generation. 
For stronger interactions, the system returns to localized behavior, as observed in Figures~\ref{fig:T_U} and \ref{fig:T_N_U}. 
The presence of third-nearest-neighbor hopping favors channel opening at $E=0$ and shifts the transmission suppression threshold to slightly lower values of $U$.

Our main results highlight that quasiperiodic geometry and electronic correlations compete, leading to a sequence of localized $\to$ delocalized $\to$ localized regimes as $U$ varies, which is in qualitative agreement with findings for two-electron wavepackets in the Anderson-Hubbard model [\onlinecite{DIAS201435}]. \textcolor{black}{The microscopic origin of this sequence is evidenced by the LDOS analysis (Fig.~\ref{fig:ldos}), which distinguishes two spatially distinct localization mechanisms: one driven by quasiperiodic geometry and the other by strong electronic repulsion.} 
This suggests a broader relevance for the interplay between interaction and localization beyond purely disordered systems.
\textcolor{black}{The oscillatory behavior observed in $T(U)$ reflects interaction-dependent modulations that mark the boundaries of the delocalized regime, and} their multiplicity increases systematically with the Fibonacci generation. 
Furthermore, the $S_3$ generation exhibits enhanced robustness against localization, due to its balanced alternation between A and B extensions when stacked, effectively introducing a periodicity that facilitates channel coupling.

Looking forward, immediate research directions include relaxing the spin-symmetric assumption to investigate \textcolor{black}{spin-dependent effects. 
Based on the established physics of zigzag edges [\onlinecite{fujita1996peculiar}], the engineered edge extensions are expected to host spin-polarized states [\onlinecite{https://doi.org/10.1002/adma.202306311}], and a fully spin-resolved treatment could reveal spin-filtering effects or spin-dependent transport enhancement modulated by the quasiperiodic sequence. 
The unique alternating pattern of the $S_3$ generation, for instance, might favor specific magnetic configurations. Such analysis, while computationally demanding for the system sizes studied here, represents a promising direction for future investigation. 
Additionally, we also intend to investigate the robustness of the observed oscillatory transmission features against structural disorder.}

From a practical perspective, our findings indicate that engineering the edge extension sequence, combined with environmental control of electronic screening, provides a promising pathway for predictably modulating transport in GNRs. This approach is especially relevant given recent advances in bottom-up fabrication techniques that achieve atomic precision in edge engineering.

\section*{Acknowledgements}

DBF acknowledges a scholarship from Coordenação de Aperfeiçoamento de Pessoal de Nível Superior (CAPES).
ALRB acknowledges financial support from CNPq (Grants 302502/2025-4 and 406836/2022-1 INCT of Spintronics and Advanced Magnetic Nanostructures - SpinNanoMag).
LFCP acknowledges financial support from CAPES (Grant 0041/2022), CNPq (Grants 436859/2018-1, 313462/2020-8,  200296/2023-0, 371610/2023-0 INCT Materials Informatics, and 310262/2025-9), FACEPE (Grant APQ-1117-1.05/22), and FINEP (Grant 0165/21).

\section{Appendix}

The Appendix describes the procedure to analyze and measure multifractal characteristics of fluctuating time series proposed in Ref. [\onlinecite{kantelhardt2002multifractal}], i.e., Multifractal Detrended Fluctuation Analysis. Starting from the time series $x_k$, we construct the series profile by performing a cumulative sum
\begin{equation}
Y(i) = \sum_{k = 1}^{i}[x_k - \langle x\rangle],\qquad i = 1,...,N
\end{equation}
where $\langle x\rangle = [\sum_{k=1}^{N}x_k]/N$ denotes the usual mean of the series. Then the profile is divided into $N_s$ non-overlapping intervals of size $s$. Within these j intervals, where $j = 1, ..., N_s$, a linear function $y_j(i) = a_j + b_j i$ is fitted to the corresponding segment of Y(i). Thus, it is possible to calculate the variance in each interval window as
\begin{equation}
F_s^2(j) = \frac{1}{s}\sum_{i=1}^{s}\{Y[(j-1)s + i] - y_j(i)\}^2.
\end{equation}
Then,  we calculate the q-th order fluctuation function as

\begin{equation}
F_q(s)= \left(\frac{1}{N_s}\sum_{j=1}^{N_s}\left[F^2_s(j)\right]^{q/2}\right)^{1/q}. \label{Fp}
\end{equation}
Once we have a set of functions $F_q(S)$, we determine the multifractal scaling exponent by looking at its relationship to the size $s$ of the interval

\begin{equation} 
F_q(s) \sim s^{h(q)},
\end{equation}
where $h(q)$ is the generalized Hurst exponent. If $h(q)$ is q-dependent, then the corresponding time series is said to be multifractal, while if h(q) is independent of q, the time series is monofractal.

The scaling exponents are further characterized through the singularity spectrum $f(\alpha)$, which provides a geometric description of the multifractal structure. It is defined as $f(\alpha)=\alpha q -\tau(q)$, where $\tau(q)=qh(q)-1$ and $\alpha = d\tau/dq$. A broad $f(\alpha)$ spectrum is characteristic of a multifractal time series, while monofractal series present a narrow $f(\alpha)$.
For the analysis of the transmission $T(E)$ time series, all series have 10,000 time steps, and we use interval sizes $s$ ranging from $400$ to $3000$.

\begin{figure}
    \centering
    \includegraphics[width=\linewidth]{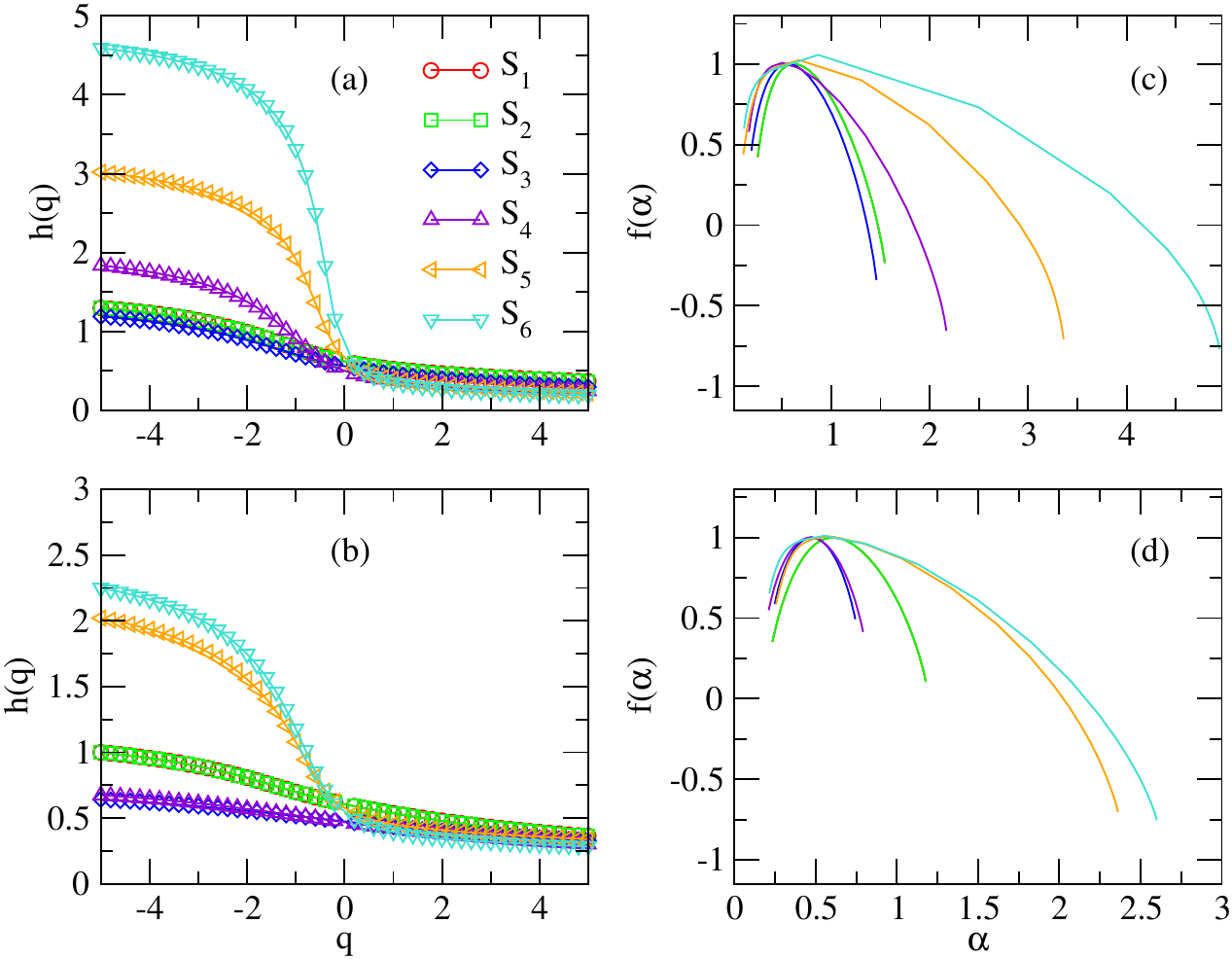}
    \caption{Multifractal analysis of charge transmission $T(E)$ time series of FEE-ZGNRs shown in Figure \ref{fig:redes_fibo}. (a,b) Generalized Hurst exponent $h(q)$ as a function of $q$ and (c,d) multifractal singularity spectrum $f(\alpha$) as a function of $\alpha$ obtained from charge transmission time series. Results in panels (a,c) were obtained considering only first-nearest-neighbor hopping  ($t_3 = 0$), while panels (b,d) include both first- and third-nearest-neighbor ($t_3 = t_1/10$).}
    \label{fig:Multifractal}
\end{figure}

Figure \ref{fig:Multifractal} shows the multifractal analysis of charge transmission $T(E)$ time series from Figure \ref{fig:T}(a) and \ref{fig:T}(b). Figure \ref{fig:Multifractal}(a) and \ref{fig:Multifractal}(b) show the generalized Hurst exponent $h(q)$ as a function of $q$ while \ref{fig:Multifractal}(c) and \ref{fig:Multifractal}(c) show the multifractal singularity spectra $f(\alpha)$ as functions of $\alpha$ obtained from the charge transmission time series. For the case considering only first neighbors ($t_3 = 0$), $h(q)$ varies with $q$ for all time series and each generation of FEE-ZGNRs, indicating a multifractality for all time series, the same is observed for the case considering first and third neighbors ($t_3 = t_1/10$). This behavior is consistent with the quasiperiodic criticality expected in Fibonacci chains [\onlinecite{10.21468/SciPostPhys.6.4.050, PhysRevB.93.205153, PhysRevResearch.3.033257, PhysRevA.35.1467, PhysRevB.35.1020, PhysRevB.34.2041}]. 

This result is supported by Fig. \ref{fig:Multifractal}(c) and \ref{fig:Multifractal}(d), which show that the $f(\alpha)$ curves are broad for all cases. 
For the case with only first-nearest-neighbors ($t_3 = 0$), the spectrum broadens with increasing generation order. 
On the other hand, in the case with first- and third-nearest-neighbors ($t_3 = t_1/10$), a similar broadening with generation  is observed, with $S_3$ and $S_4$ exhibiting the least broad spectra.

\bibliographystyle{apsrev4-1}
\bibliography{ref.bib}


\end{document}